\documentclass[11pt]{article}
\pdfoutput=1
\usepackage{jheppub}

\begin{document}


\title{\boldmath Cardy-like asymptotics of the 4d $\mathcal{N}=4$ index and AdS$_5$ blackholes}

\author{Arash Arabi Ardehali}

\affiliation{Department of Physics and Astronomy, Uppsala University,\\
Box 516, SE-751 20 Uppsala, Sweden}

\affiliation{\&\\
School of Physics, Institute
for Research in Fundamental Sciences (IPM),\\
P.O.Box 19395-5531, Tehran, Iran}

\emailAdd{ardehali@physics.uu.se}

\abstract{Choi, Kim, Kim, and Nahmgoong have recently pioneered
analyzing a Cardy-like limit of the superconformal index of the 4d
$\mathcal{N}=4$ theory with complexified fugacities which encodes
the entropy of the dual supersymmetric AdS$_5$ blackholes. Here we
study the Cardy-like asymptotics of the index within the rigorous
framework of elliptic hypergeometric integrals, thereby filling a
gap in their derivation of the blackhole entropy function, finding a
new blackhole saddle-point, and demonstrating novel bifurcation
phenomena in the asymptotics of the index as a function of fugacity
phases. We also comment on the relevance of the supersymmetric
Casimir energy to the blackhole entropy function in the present
context.}

\maketitle \flushbottom

\section{Introduction}

It has been a long-standing challenge in AdS$_5$/CFT$_4$ to
reproduce the entropy of the charged, rotating, BPS, asymptotically
AdS$_5$ blackholes of
\cite{GR:2004a,GR:2004b,KLR:2006,CCLP:2005,CGLP:2005b} from a
microscopic counting of BPS states in the 4d $\mathcal{N}=4$ CFT.
Several attempts in this direction were made in the past fifteen
years or so (\emph{e.g.}
\cite{Kinney:2005ej,Berkooz:2006,Grant:2008,Berkooz:2008,Chang:2014}),
leading to various new lessons for holography and superconformal
field theory (SCFT), but not to the desired microscopic count.

In particular, an index was devised in
\cite{Kinney:2005ej,Romelsberger:2005eg} for counting the BPS states
of general 4d SCFTs. The index counts all of the states---in the
radial quantization of the SCFT---that are annihilated by a chosen
supercharge. We adopt conventions in which such states satisfy the
``BPS condition'' $\Delta-J_1-J_2-\frac{3}{2}r=0$, where
$\Delta,J_1,J_2,r$ are the quantum numbers of the 4d $\mathcal{N}=1$
superconformal group SU($2,2|1$). The index
\begin{equation}
    \begin{split}
        \mathcal{I}(p,q,u_{k}):=\mathrm{Tr}\left[(-1)^F e^{\hat{\beta}(\Delta-J_1-J_2-\frac{3}{2}r)} p^{J_1+\frac{r}{2}}q^{J_2+\frac{r}{2}}\prod_k u_k^{q_k}\right],
    \end{split}\label{eq:indexDef}
\end{equation}
is thus independent of $\hat{\beta}$, but it does depend on the
\emph{spacetime fugacities} $p,q$, as well as the \emph{flavor
fugacities} $u_k$ associated with flavor quantum numbers $q_k$
commuting with the supercharge. In the case of the $\mathcal{N}=4$
theory, the SU(4) $R$-symmetry of the $\mathcal{N}=4$ superconformal
algebra decomposes into SU($3)\times$U(1)$_{r}$, so there is an
SU(3) ``flavor'' symmetry group commuting with the chosen
supercharge; hence there are three $q_k$ with $\sum_{k=1}^3 q_k=0$,
and three $u_k$ satisfying $\prod_{k=1}^3 u_k=1$. It is customary to
define $y_k:=(pq)^{1/3}u_k$ and $Q_k:=q_k+r/2$. Then, dismissing
$\hat{\beta}$, we can rewrite the index of the $\mathcal{N}=4$
theory as
\begin{equation}
    \begin{split}
        \mathcal{I}(p,q,y_{1,2,3})=\mathrm{Tr}\left[(-1)^F
        p^{J_1}q^{J_2}y_1^{Q_1}y_2^{Q_2}y_3^{Q_3}\right].
    \end{split}\label{eq:indexDefN=4}
\end{equation}
This index was computed at finite rank for the U($N$)
$\mathcal{N}=4$ theory in the original paper \cite{Kinney:2005ej}.
Then, in an initial attempt to make contact with holography, the
large-$N$ limit of the index was evaluated for \emph{real-valued
fugacities} and was seen to be $O(N^0)$; the result perfectly
matched the index of the KK supergravity multi-particle states in
the dual AdS$_5$ theory, but clearly could not account for the
$O(N^2)$ entropy of the bulk supersymmetric AdS$_5$ blackholes
\cite{Kinney:2005ej}. For some time this negative result was
interpreted as an indication that the index does not encode the bulk
blackhole microstates.

Very recently it has been discovered by Choi, Kim, Kim, and
Nahmgoong (CKKN) \cite{Choi:2018b}, and independently by Benini and
Milan \cite{Benini:2018}, that allowing the five fugacities in the
index to take complex values one can achieve the desired
$O(e^{N^2})$ behavior in the large-$N$ limit of the index. Benini
and Milan have succeeded in directly obtaining the AdS$_5$ blackhole
entropy function in the large-$N$ limit of the index
\cite{Benini:2018}, while CKKN took a different route and derived
the entropy function in a double-scaling---Cardy-like as well as
large-$N$---limit \cite{Choi:2018b,Choi:2018}. In the present paper
we derive the entropy function in a Cardy-like limit of the index at
finite rank; although our analysis is closely related to that of
CKKN \cite{Choi:2018}, ours is more analogous to the Cardy-formula
\cite{Cardy:1986ie} derivations of blackhole entropy in
AdS$_3$/CFT$_2$ (\emph{e.g.}
\cite{Strominger:1997nh,Strominger:1996,Breckenridge:1997}) where
the central charge is kept fixed.\\

The study of the Cardy-like asymptotics of 4d superconformal indices
had some history prior to \cite{Choi:2018}, but was again mostly
limited to real-valued fugacities (\emph{e.g.}
\cite{Ardehali:2014,DiPietro:2014,Ardehali:2015b,Ardehali:2015c,Ardehali:thesis,DiPietro:2017,Hwang:2018}).
The idea that blackhole microstate counting requires complex-valued
fugacities in the $\mathcal{N}=4$ index was not properly appreciated
until the recent work of Hosseini, Hristov, and Zaffaroni (HHZ)
\cite{Hosseini:2017}. This work provided the impetus for the later
investigations of CKKN \cite{Choi:2018b,Choi:2018} and Benini-Milan
\cite{Benini:2018}. HHZ started from the supergravity side and
bridged half-way towards the CFT by presenting a ``grand-canonical''
function---henceforth the HHZ function---from which a Legendre
transform gives the micro-canonical entropy of the AdS$_5$
blackholes; the remaining challenge was to extract the HHZ function
in an appropriate asymptotic regime from the index. In particular,
it was understood by HHZ \cite{Hosseini:2017} (based on recent
lessons from AdS$_4$/CFT$_3$ \cite{BHZ:2016,BHZ:2017}) that
complexified fugacities are needed in the index in order to make
contact with the grand-canonical function of the AdS$_5$ blackholes.
As alluded to above, CKKN \cite{Choi:2018} and independently Benini
and Milan \cite{Benini:2018} have recently completed the bridge
between the CFT and the bulk by deriving the HHZ function through
asymptotic analysis of the $\mathcal{N}=4$ theory index, the first
group in a double-scaling limit and the second group in a large-$N$
limit.

In the present paper we analyze the Cardy-like asymptotics of the
$\mathcal{N}=4$ theory index with complexified fugacities using the
rigorous machinery of elliptic hypergeometric integrals
\cite{vanD:2001,Spiridonov:2003,Dolan:2008,Rains:2009} in various
Cardy-like regimes of parameters where the flavor fugacities
approach the unit circle and the spacetime fugacities approach $1$.
In particular, we fill a gap in the CKKN derivation of the HHZ
function in this limit by showing that the eigenvalue configuration
they chose in their asymptotic analysis of the matrix-integral
expression for the index is indeed the dominant configuration in the
regime of parameters pertaining to the blackhole saddle-point they
considered (though we demonstrate that it in fact fails to be the
dominant configuration in essentially half of the parameter-space,
where a more intricate analysis is called for). Moreover, we
discover a new blackhole saddle-point in a different regime of
parameters, corresponding to fugacities that are complex conjugate
to those at the CKKN saddle-point. We present intuitive arguments
suggesting that no other blackhole saddle-points with such large
entropies exist in the Cardy-like limit. We also illustrate
interesting dependence of the qualitative behavior of the Cardy-like
asymptotics of the index on the complex phases of the
fugacities.\\

In the rest of this introduction we give a sketchy account of the
asymptotic analysis extracting the blackhole entropy function from
the appropriate Cardy-like limit of the superconformal index of the
$\mathcal{N}=4$ theory. The main body of the paper starts in
Section~\ref{sec:betaComplex} where we elaborate on the sketchy
derivation of the present section; we study the Cardy-like
asymptotics of the $\mathcal{N}=4$ theory index with all its
fugacities complexified, clarifying---and addressing a gap in---the
CKKN derivation of the HHZ function. A thorough enough understanding
of the asymptotics of the index in different Cardy-like regimes of
parameters results in that section which reveals a second blackhole
saddle-point in a regime complementary to that of CKKN, and moreover
allows us to argue intuitively that no further saddle-points with
such large entropies exist. In Section~\ref{sec:betaReal} we keep
the spacetime fugacities real-valued, and demonstrate novel
bifurcation phenomena in the Cardy-like asymptotics of the index as
a function of the flavor-fugacity phases. Section~\ref{sec:Casimir}
discusses the relation between the Hamiltonian superconformal index
and the Lagrangian index computed through path-integration; the two
differ by a Casimir-energy factor which is argued to be irrelevant
to the blackhole entropy function in the present context. Finally,
Section~\ref{sec:open} discusses the important open ends of the
present work.

\subsection{Outline of the CKKN
derivation in the elliptic hypergeometric
language}\label{subsec:outline}

We now present an outline of the CKKN derivation \cite{Choi:2018} of
the HHZ function \cite{Hosseini:2017}, translated to the language of
elliptic hypergeometric integrals. More precisely, the problem we
consider differs from that of \cite{Choi:2018} in two respects:
\begin{itemize}
\item while \cite{Choi:2018} considered the $\mathcal{N}=4$ theory with U($N$)
gauge group, we consider the SU($N$) theory---the details are rather
similar and the end results are related via $N^2\to N^2-1$ shifts;
\item while in \cite{Choi:2018} a double-scaling---Cardy-like as well as
large-$N$---limit is taken to simplify the analysis, here in analogy
with the Cardy-formula derivations of blackhole entropy in
AdS$_3$/CFT$_2$ we keep $N$ finite and only take a Cardy-like limit.
\end{itemize}

\subsection*{The special function as the starting point}

The superconformal index of the SU($N$) $\mathcal{N}=4$ theory is
given by the following elliptic hypergeometric integral (see e.g.
\cite{Spiridonov:2010sv}):
\begin{equation}
    \begin{split}
        \mathcal{I}(p,q,y_{1,2,3}):&=\mathrm{Tr}\left[(-1)^F p^{J_1}q^{J_2}y_1^{Q_1}y_2^{Q_2}y_3^{Q_3}\right]\\
        &=\frac{\big((p;p)(q;q)\big)^{N-1}}{N!}\prod_{k=1}^{3}\Gamma^{N-1}\big(y_k\big)\oint\prod_{j=1}^{N-1}\frac{\mathrm{d}z_j}{2\pi i z_j}\ \prod_{1\le i<j\le N}\frac{\prod_{k=1}^{3}\Gamma\big(y_k (z_i/z_j)^{\pm 1}\big)}{\Gamma\big( (z_i/z_j)^{\pm 1}\big)},
    \end{split}\label{eq:EHI}
\end{equation}
with the unit-circle contour for the $z_j=e^{2\pi i x_j}$ while
$\prod_{j=1}^N z_j=1$, and with $p,q,y_k$ strictly inside the unit
circle such that $\prod_{k=1}^3 y_k=pq$. The two special functions
$(\cdot;\cdot)$ and $\Gamma(\cdot)$ are respectively the
\emph{Pochhammer symbol} and the \emph{elliptic gamma function}
\cite{Ruijsenaars:1997}:
\begin{equation}
(a;q):=\prod_{k=0}^{\infty}(1-a q^k),\label{eq:PochDef}
\end{equation}
\begin{equation}
\Gamma(z):=\prod_{j,k\ge 0}\frac{1-z^{-1}p^{j+1}q^{k+1}}{1-z
p^{j}q^{k}},\label{eq:GammaDef}
\end{equation}
and $\Gamma(z^{\pm1})$ stands for $\Gamma(z)\Gamma(z^{-1})$.

The integral expression gives the index as a meromorphic function of
$p,q,y_k$ in the domain $0<|p|,|q|,|y_k|<1$. A contour deformation
can presumably allow meromorphic continuation of the index to
$0<|p|,|q|<1$, $y_k\in\mathbb{C}^\ast$ (\emph{c.f.}
\cite{Rains:2005}).

\subsection*{Asymptotic analysis in the limit encoding blackholes}

The Cardy-type limit analyzed prior to the work of CKKN
\cite{Choi:2018} was of the form $p,q,y_k\to 1$; more precisely, it
was what in the mathematics literature is referred to as the
\emph{hyperbolic limit} of the elliptic hypergeometric integral
\cite{Rains:2009,Ardehali:2018a}. CKKN considered instead limits of
the type $p,q\to 1$, $y_{i}\to e^{i\theta_i}$, with $\theta_i\notin
2\pi\mathbb{Z}$: they correctly recognized that giving finite
(non-vanishing) phases to the flavor fugacities can obstruct the
bose-fermi cancelations\footnote{A similar obstruction mechanism is
at work in the AdS$_3$/CFT$_2$ context, where the entropy of the
AdS$_3$ blackholes is derived from a Cardy-like limit of the CFT$_2$
elliptic genus $\chi(q,y)$: the limit $q,y\to1$ does not encode the
bulk blackholes, but the limit $q\to1$, $y\to e^{i\theta}$ with
$\theta\notin 2\pi\mathbb{Z}$ does. However, note that while in the
AdS$_3$/CFT$_2$ context $q$ can be kept real, in AdS$_5$/CFT$_4$ the
spacetime fugacities $p,q$ should take off the real line to meet the
blackhole saddle-points. See \cite{Benini:2018,BHZ:2016,BHZ:2017}
for related discussions of ``$\mathcal{I}$-extremization'' in the
large-$N$ analysis.} occurring in the hyperbolic limit. For future
reference we define $\sigma,\tau,\Delta_k$ through $p=e^{2\pi i
\sigma},q=e^{2\pi i\tau}, y_k=e^{2\pi i \Delta_k}$, and write the
appropriate limit explicitly as
\begin{equation*}
\text{\emph{the CKKN limit:}}\quad
|\sigma|,|\tau|,\mathrm{Im}\Delta_k\to 0,\ \text{with\
}\frac{\tau}{\sigma}\in\mathbb{R}_{>0},\mathrm{Re}\Delta_{k}\
\text{fixed},\ \text{and}\ \mathrm{Im}\tau,\mathrm{Im}\sigma>0.
\end{equation*}
Note that the ``balancing condition'' $\prod_{k=1}^3 y_k=pq$ implies
$\sum_{k=1}^3 \Delta_k-\sigma-\tau\in\mathbb{Z}$, and that the
restriction $\mathrm{Im}\tau,\mathrm{Im}\sigma>0$ keeps us in the
domain of meromorphy of the index.

The asymptotic analysis of the integral (\ref{eq:EHI}) now proceeds
as follows. As will be explained in Section~\ref{sec:betaComplex},
the leading asymptotics comes from the elliptic gamma functions
$\Gamma(\cdot)$, so the Pochhammer symbols $(\cdot;\cdot)$ and the
$N!$ in the pre-factor can be neglected. The required estimate,
reviewed in Section~\ref{sec:betaComplex}, follows from
Proposition~2.11 of Rains \cite{Rains:2009}:
\begin{equation}
    \Gamma(e^{2\pi i x})= e^{-2\pi i\frac{\kappa(x)}{12\tau\sigma}+\mathcal{O}(\frac{\tau+\sigma}{\tau\sigma})},\label{eq:leadingEstimate}
\end{equation}
for $|\tau|,|\sigma|\to0$, with
$\mathrm{Im}\tau,\mathrm{Im}\sigma>0$, and
$\frac{\tau}{\sigma}\in\mathbb{R}_{>0}$,
$x\in\mathbb{R}/\mathbb{Z}$. Here $\kappa(\cdot)$ is the continuous,
odd, piecewise cubic\footnote{Hence the ``k''appa symbol introduced
for it in \cite{Ardehali:2015c}.}, periodic function
\begin{equation}
\begin{split}
\kappa(x)&:=\{x\}(1-\{x\})(1-2\{x\})\\
&\left(=2x^3-3x|x|+x \quad\quad\text{for
$x\in[-1,1]$}\right),\label{eq:kappaDef}
\end{split}
\end{equation}
with $\{x\}=x-\lfloor x\rfloor$.

In order to apply the estimate (\ref{eq:leadingEstimate}) to the
gamma functions in (\ref{eq:EHI}) we have to identify the phase of
the arguments with $2\pi x$; then, for instance, we can apply
(\ref{eq:leadingEstimate}) to the gamma function in the numerator of
the integrand of (\ref{eq:EHI}) by identifying $x$ with
$\mathrm{Re}\Delta_k\pm(x_i-x_j)$. This way we can simplify
(\ref{eq:EHI}) to
\begin{equation}
    \begin{split}
        \mathcal{I}(p,q,y_{1,2,3})\xrightarrow{\text{in the CKKN limit}}\int_{-1/2}^{1/2}\mathrm{d}^{N-1}\boldsymbol{x}\
        e^{-2\pi i\frac{Q_h(\boldsymbol{x};\mathrm{Re}\Delta_k)}{\tau\sigma}},
    \end{split}\label{eq:EHIsimplified}
\end{equation}
with $Q_h$ given by\footnote{Compare with the $Q_h$ function defined
in \cite{Ardehali:2015c}; gauge anomaly cancelation implies that
both are piecewise ``Q''uadratic as a function of $\boldsymbol{x}$.
The subscript $h$ is used because the CKKN limit is a variant of the
``h''yperbolic limit of the elliptic hypergeometric integral.}
\begin{equation}
\begin{split}
Q_h(\boldsymbol{x};\mathrm{Re}\Delta_k):=\frac{1}{12}\sum_{k=1}^3\left((N-1)\kappa(\mathrm{Re}\Delta_k)+\sum_{1\le
i<j\le N}\kappa(\mathrm{Re}\Delta_k\pm
(x_i-x_j))\right),\label{eq:QhDef}
\end{split}
\end{equation}
where $\kappa(A\pm B)$ stands for $\kappa(A+B)+\kappa(A-B)$. It only
remains to evaluate the asymptotics of the integral
(\ref{eq:EHIsimplified}).

Note that we are assuming $\mathrm{Im}(\tau\sigma)\neq0$; this
corresponds to complexifying the ``temperature'' as explained below.
When $\mathrm{Im}(\tau\sigma)=0$ the integrand of
(\ref{eq:EHIsimplified})---or already the RHS of
(\ref{eq:leadingEstimate})---would be a pure phase, and not
sufficient to describe the exponential growth of the blackhole
microstates. The $\mathrm{Im}(\tau\sigma)=0$ case is therefore not
directly relevant to the AdS$_5$ blackhole physics, but it exhibits
some interesting asymptotic bifurcation phenomena that are discussed
in Section~\ref{sec:betaReal}.

The last step of the asymptotic analysis of the index involves
arguing that in the appropriate range of parameters the dominant
small-$|\tau|,|\sigma|$ configurations in (\ref{eq:EHIsimplified})
have $x_i-x_j=0$ (implying $x_j=\mathrm{const}$, which in our
SU($N$) case would mean that all the holonomies are equal to
$\frac{n}{N}$ for some $n\in\{0,1,\dots,N-1\}$). CKKN simply assumed
\cite{Choi:2018} this to be the case. In
Section~\ref{sec:betaComplex} we will prove that for the range of
parameters relevant to the AdS$_5$ blackholes (\emph{e.g.} for
$\mathrm{Im}(\tau\sigma)>0$ \&
$\mathrm{Re}\Delta_{1,2},-1-\mathrm{Re}\Delta_{1}-\mathrm{Re}\Delta_{2}\in]-1,0[$)
their assumption is correct. Hence the asymptotics of the index
becomes\footnote{Compare with Eq.~(2.34) of CKKN \cite{Choi:2018};
note that $2\pi i
\Delta^{\mathrm{here}}_k=-\Delta_k^{\mathrm{CKKN}}$, while $2\pi
i\{\sigma,\tau\}=-\omega_{\{1,2\}}^{\mathrm{CKKN}}$.}
\begin{equation}
    \begin{split}
        \log\mathcal{I}(p,q,y_{1,2,3})\approx-\frac{2\pi i}{\tau\sigma}Q_h(0;\mathrm{Re}\Delta_k)=
        -2\pi i\frac{N^2-1}{12\tau\sigma}\sum_{k=1}^3
        \kappa(\mathrm{Re}\Delta_k).
    \end{split}\label{eq:EHIasy1}
\end{equation}
The right-hand-side is a nonanalytic function of the
$\mathrm{Re}\Delta_k$, manifestly invariant under
$\mathrm{Re}\Delta_k\to \mathrm{Re}\Delta_k+1$ as it should be.

To match the grand-canonical function of HHZ \cite{Hosseini:2017} we
now pick a particular chamber in the parameter-space so that an
analytic expression can be written down. Specifically, assuming
$\mathrm{Im}(\tau\sigma)>0$, going into the chamber
$-1<\mathrm{Re}\Delta_{1,2,3}<0$ with
$\mathrm{Re}\Delta_3=-1-\mathrm{Re}\Delta_1-\mathrm{Re}\Delta_2$, we
can simplify $\sum_{k=1}^3 \kappa(\mathrm{Re}\Delta_k)$ to
$6\mathrm{Re}\Delta_{1}\mathrm{Re}\Delta_{2}\mathrm{Re}\Delta_{3}$,
and arrive at
\begin{equation}
    \begin{split}
        \log\mathcal{I}(p,q,y_{1,2,3})\approx
        -2\pi
        i\frac{N^2-1}{2\tau\sigma}\mathrm{Re}\Delta_{1}\mathrm{Re}\Delta_{2}\mathrm{Re}\Delta_{3}.
    \end{split}\label{eq:EHIasy2}
\end{equation}
Analytic continuation of (\ref{eq:EHIasy2}) to complex $\Delta_k$
(\emph{i.e.} replacing every $\mathrm{Re}\Delta_{k}$ with
$\Delta_k$) allows recovering the subleading terms in the CKKN limit
and connecting with the complex HHZ function \cite{Hosseini:2017}
(see the end of Section~\ref{sec:betaComplex} for more details):
\begin{equation}
    \begin{split}
        \log\mathcal{I}(p,q,y_{1,2,3})\approx
        -2\pi
        i\frac{N^2-1}{2\tau\sigma}\Delta_{1}\Delta_{2}\Delta_{3},
    \end{split}\label{eq:EHIasyHHZ}
\end{equation}
with $\Delta_{1}+\Delta_{2}+\Delta_{3}-\tau-\sigma=-1$.\\

So far in this subsection we have been essentially rephrasing the
developments due to CKKN \cite{Choi:2018}. One of the novel
contributions of the present paper is to demonstrate in
Section~\ref{sec:betaComplex} that when $\mathrm{Im}(\tau\sigma)<0$
another chamber with $0<\mathrm{Re}\Delta_{1,2,3}<1$ and
$\mathrm{Re}\Delta_3=1-\mathrm{Re}\Delta_1-\mathrm{Re}\Delta_2$
yields the asymptotics (\ref{eq:EHIasyHHZ}) in the CKKN limit, this
time with $\Delta_{1}+\Delta_{2}+\Delta_{3}-\tau-\sigma=+1$.

\subsection*{Legendre transform and blackhole entropy}

Thinking of the index (\ref{eq:indexDefN=4}) as the generating
function of the degeneracies $d(J_{1,2},Q_{1,2,3})$ of the BPS
states\footnote{Although at first glance it appears that because of
the $(-1)^F$ factor in it the index (\ref{eq:indexDefN=4}) counts
the number of bosonic states minus the number of fermionic states,
as argued in \cite{Benini:2018}, on the blackhole saddle-points
essentially all the states are expected to be bosonic, so the index
counts a degeneracy.} in the $\mathcal{N}=4$ theory, methods of
elementary analytic combinatorics can be used to extract the
large-$J_{1,2},Q_{1,2,3}$ asymptotics of $d(J_{1,2},Q_{1,2,3})$ from
the Cardy-like asymptotics of the index. The CKKN limit of the index
encodes the degeneracy of the BPS states as $Q_{1,2,3}\sim \Lambda,\
J_{1,2}\sim\Lambda^{3/2},\ \Lambda\to\infty$, such that the charge
relation \cite{Choi:2018,Cabo-Bizet:2018}
\begin{equation}
Q_1Q_2Q_3+\frac{N^2-1}{2}J_1J_2=\big(Q_1+Q_2+Q_3+\frac{N^2-1}{2}\big)\big(Q_1Q_2+Q_2Q_3+Q_3Q_1-\frac{N^2-1}{2}(J_1+J_2)\big),\label{eq:chargeRelation}
\end{equation}
of the bulk AdS$_5$ blackholes is satisfied \cite{Choi:2018}.

The degeneracies can be obtained from the generating function
through
\begin{equation}
d(J_{1,2},Q_{1,2,3})=\oint
\mathcal{I}(p,q,y_{1,2,3})p^{-J_1}q^{-J_2}(\prod_{k=1}^3
y_k^{-Q_k})\ \frac{\mathrm{d}p}{2\pi i p}\frac{\mathrm{d}q}{2\pi i
q}\left(\prod_{k=1}^2\frac{\mathrm{d}y_k}{2\pi i y_k}\right),
\end{equation}
with all of the contours slightly inside the unit circle; note that
$y_3$ is not independent, so is not integrated over on the RHS
(\emph{c.f.} Section~5 of \cite{Benini:2018}). The asymptotic
degeneracy can be obtained using a saddle-point evaluation of the
integral on the right-hand side. Using the Cardy-like asymptotics in
(\ref{eq:EHIasyHHZ}), the result for the asymptotic entropy
$S(J_{1,2},Q_{1,2,3})=\log d(J_{1,2},Q_{1,2,3})$ becomes
\begin{equation}
S(J_{1,2},Q_{1,2,3})\approx\left(-2\pi
i\frac{N^2-1}{2\tau\sigma}\Delta_1\Delta_2\Delta_3-2\pi i \sigma
J_1-2\pi i\tau J_2-\sum_{k=1}^3 2\pi i \Delta_k
Q_k\right)_{\mathrm{ext}},\label{eq:ext}
\end{equation}
with $\sum \Delta_k-\tau-\sigma=-1$, as well as
$-1<\mathrm{Re}\Delta_{1,2,3}<0$, $\mathrm{Im}\tau\sigma>0$. The
subscript ``ext'' on the RHS means picking its extremized value on
the saddle-point.

The extremization problem was addressed for the
$\mathrm{Im}\tau\sigma>0$ case by HHZ \cite{Hosseini:2017}, but was
made completely explicit and analytic by CKKN \cite{Choi:2018} (and
independently in Appendix~B of \cite{Cabo-Bizet:2018} by Cabo-Bizet,
Cassani, Martelli, and Murthy), who found the blackhole saddle-point
at
\begin{equation}
\Delta_k=-\frac{\frac{1}{S-2\pi iQ_k}}{\sum_{j=1}^3\frac{1}{S-2\pi
iQ_j}-\sum_{l=1}^2\frac{1}{S+2\pi iJ_l}},\quad
\{\sigma,\tau\}=-\frac{\frac{1}{S+2\pi
iJ_{\{1,2\}}}}{\sum_{j=1}^3\frac{1}{S-2\pi
iQ_j}-\sum_{l=1}^2\frac{1}{S+2\pi iJ_l}},\label{eq:CKKNHHZsaddle}
\end{equation}
satisfying $\Delta_{1}+\Delta_{2}+\Delta_{3}-\tau-\sigma=-1$, and
giving the entropy
\begin{equation}
S\approx
2\pi\sqrt{Q_1Q_2+Q_2Q_3+Q_3Q_1-\frac{N^2-1}{2}(J_1+J_2)},\label{eq:entropy1}
\end{equation}
which thanks to the charge relation (\ref{eq:chargeRelation}) can be
written in the alternative form
\begin{equation}
S\approx
2\pi\sqrt{\frac{Q_1Q_2Q_3+\frac{N^2-1}{2}J_1J_2}{Q_1+Q_2+Q_3+\frac{N^2-1}{2}}}.\label{eq:entropy2}
\end{equation}
Both of the relations (\ref{eq:entropy1}), (\ref{eq:entropy2})
correctly reproduce the Bekenstein-Hawking entropy of the BPS
AdS$_5$ blackholes of
\cite{GR:2004a,GR:2004b,KLR:2006,CCLP:2005,CGLP:2005b}, upon using
the AdS/CFT dictionary
$N^2-1=\frac{\pi\ell_{\mathrm{AdS}_5}}{2G_{\mathrm{AdS}_5}}$, with
$\ell_{\mathrm{AdS}_5},G_{\mathrm{AdS}_5}$ respectively the radius
and the Newton constant of the bulk $\mathrm{AdS}_5$.\\

Note that the term $\frac{N^2-1}{2}(J_1+J_2)$ in (\ref{eq:entropy1})
and the term $\frac{N^2-1}{2}$ in (\ref{eq:entropy2}) are subleading
in the CKKN scaling limit $J^2\sim Q^3\to\infty$; capturing them was
the sole purpose of keeping the subleading terms in
(\ref{eq:EHIasyHHZ}) (compared to (\ref{eq:EHIasy2})), as well as in
(\ref{eq:chargeRelation}), (\ref{eq:ext}), and
(\ref{eq:CKKNHHZsaddle}).\\

In Section~\ref{sec:betaComplex} we show that (\ref{eq:EHIasyHHZ})
is valid in the CKKN limit also when
$0<\mathrm{Re}\Delta_{1,2,3}<1$, $\mathrm{Im}\tau\sigma<0$, though
this time with $\Delta_{1}+\Delta_{2}+\Delta_{3}-\tau-\sigma=+1$,
and find a new blackhole saddle-point at
\begin{equation}
\Delta_k=\frac{\frac{1}{S+2\pi iQ_k}}{\sum_{j=1}^3\frac{1}{S+2\pi
iQ_j}-\sum_{l=1}^2\frac{1}{S-2\pi iJ_l}},\quad
\{\sigma,\tau\}=\frac{\frac{1}{S-2\pi
iJ_{\{1,2\}}}}{\sum_{j=1}^3\frac{1}{S+2\pi
iQ_j}-\sum_{l=1}^2\frac{1}{S-2\pi iJ_l}},\label{eq:NEWsaddle}
\end{equation}
with the same entropy $S$ as that of the CKKN
saddle-point\footnote{The observation that the HHZ function with
$\Delta_{1}+\Delta_{2}+\Delta_{3}-\tau-\sigma=+1$ also happens to
give the right entropy has been made previously in Appendix~B of
\cite{Cabo-Bizet:2018}. Our contribution here is to clarify the
physics of this observation by showing that the HHZ function with
$\Delta_{1}+\Delta_{2}+\Delta_{3}-\tau-\sigma=+1$ actually arises as
the asymptotics of the index in a regime of parameters separate from
that considered by CKKN.}. We moreover argue that besides the two
just described---having complex conjugate fugacities
$p,q,y_{1,2,3}$---no other blackhole saddle-points with such large
entropies exist in the Cardy-like asymptotics of the $\mathcal{N}=4$
theory index.

\subsection*{Final remarks}

\textbf{A remaining gap for unequal} $\mathbf{Q_k}$. A rather
serious gap in the above derivation is revealed upon closer
inspection of the critical $\Delta_k$ in (\ref{eq:CKKNHHZsaddle})
and (\ref{eq:NEWsaddle}): while our asymptotic analysis is valid
only in the limit $\mathrm{Im}\Delta_k\to0,$ the blackhole
saddle-points have nonzero $\mathrm{Im}\Delta_k$ unless
$Q_1=Q_2=Q_3$. It is therefore only in the special case with
equal---or approximately equal---charges that the above derivation
(augmented with the refinements of Section~\ref{sec:betaComplex}) is
satisfactory. CKKN assumed in a leap of faith \cite{Choi:2018} that
the asymptotics (\ref{eq:EHIasyHHZ}) remains valid away from the
limit $\mathrm{Im}\Delta_k\to0,$ and thus the blackhole entropy
derivation can be extended to the general case with unequal charges.
In Section~\ref{sec:betaComplex} we present a partial justification
for this extrapolation; the complete justification is beyond the
scope of the present paper, and its absence constitutes the most
important open end of this work.\\

\noindent\textbf{Cardy-like versus large-}$\mathbf{N}$. The above
derivation extracts the AdS$_5$ blackhole entropy from a
``high-temperature'' (Cardy-like) limit of the 4d superconformal
index at finite $N$. This is analogous to how the classic papers of
Strominger-Vafa \cite{Strominger:1996}, BMPV
\cite{Breckenridge:1997}, and Strominger \cite{Strominger:1997nh}
derived the Bekenstein-Hawking entropy of certain blackholes in what
nowadays might be called an AdS$_3$/CFT$_2$ context.

From the holographic perspective, a more conceptually satisfying
derivation would involve the large-$N$ limit of the index. In
AdS$_3$/CFT$_2$ such conceptually satisfactory derivations can be
found in \cite{Sen:2011,Hartman:2014}. In the AdS$_5$/CFT$_4$
context this was achieved very recently by Benini and Milan
\cite{Benini:2018}, leveraging the Bethe Ansatz formula of Closset,
Kim, and Willett \cite{Closset:2017Bethe}. Curiously, although the
derivation in \cite{Benini:2018} is not limited to the equal-charge
blackholes, because of certain technical obstacles it so far applies
only to the case with equal angular momenta $J_1= J_2$ and the
general case with $J_1\neq J_2$ is still open. The more general
Bethe Ansatz formula of \cite{Benini:2018Bethe} seems promising in
that direction.

%
%
%
\section{Complexified temperature and AdS$_5$ blackholes
($|\beta|\to 0$,
$0<|\mathrm{arg}\beta|<\frac{\pi}{2}$)}\label{sec:betaComplex}

In this section we fill in the gaps of
Subsection~\ref{subsec:outline}. In particular, we give a rigorous
derivation of the estimate (\ref{eq:leadingEstimate}) for the
elliptic gamma function, fill the gap in the CKKN asymptotic
analysis in the region $\mathrm{Im}\tau\sigma>0$, extend the
analysis to the region $\mathrm{Im}\tau\sigma<0$ where we find a new
blackhole saddle-point, and argue that no further blackhole
saddle-points with such large entropies exist.

\subsection*{The elliptic gamma function estimate (\ref{eq:leadingEstimate})}

Let us define the parameters $b,\beta$ through $\tau=\frac{i\beta
b^{-1}}{2\pi}$, $\sigma=\frac{i\beta b}{2\pi}$. For
$p,q\in\mathbb{R}$, the parameter $\beta$ defined as such was
referred to as the \emph{inverse-temperature} in
\cite{Ardehali:2015b,Ardehali:2015c}; here we similarly refer to
$\beta$ as the \emph{complexified inverse-temperature}. Throughout
the present work we assume $b\in\mathbb{R}_{>0}$ (\emph{i.e.}
$\tau/\sigma\in\mathbb{R}_{>0}$); this simplifies the analysis and
suffices for making contact with blackhole physics in the Cardy-like
limit. We also take $\mathrm{Re}\beta>0$ (\emph{i.e.}
$|\mathrm{arg}\beta|<\frac{\pi}{2}$) to stay within the domain of
meromorphy of the index (\ref{eq:EHI}). In terms of $b,\beta$ we
have
\begin{equation*}
\text{\emph{the CKKN limit:}}\quad |\beta|,\mathrm{Im}\Delta_k\to
0,\ \text{with\ }b\in\mathbb{R}_{>0},\mathrm{Re}\Delta_{k}\
\text{fixed},\ \text{and}\ \mathrm{Re}\beta>0.
\end{equation*}

The starting point for deriving the estimate
(\ref{eq:leadingEstimate}) is the following identity, essentially
due to Narukawa \cite{Narukawa:2004}:
\begin{equation}
\begin{split}
\Gamma(e^{2\pi i x})=e^{2\pi i
Q_{+}(x;\sigma,\tau)}\psi_b(\text{\footnotesize{$-\frac{2\pi i
x}{\beta}-\frac{b+b^{-1}}{2}$}})\prod_{n=1}^{\infty}\frac{\psi_b(-\frac{2\pi
in}{\beta}-\frac{2\pi i
x}{\beta}-\frac{b+b^{-1}}{2})}{\psi_b(-\frac{2\pi
in}{\beta}+\frac{2\pi i
x}{\beta}+\frac{b+b^{-1}}{2})},\label{eq:GammaFVSq}
\end{split}
\end{equation}
where
\begin{equation}
\begin{split}
Q_{+}(x;\sigma,\tau)=&-\frac{x^3}{6\tau\sigma}+\frac{\tau+\sigma+1}{4\tau\sigma}x^2-\frac{\tau^2+\sigma^2+3\tau\sigma+3\tau+3\sigma+1}{12\tau\sigma}x\\
&+\frac{1}{24}(\tau+\sigma+1)(1+\tau^{-1}+\sigma^{-1}),\label{eq:QpDef}
\end{split}
\end{equation}
and $\psi_b(x)$ a function [see Appendix~A of \cite{Ardehali:2015c}
for its definition in terms of the hyperbolic gamma function] with
the important property that for $\mathrm{arg}x$ inside compact
subsets of $(-\pi,0)$, and fixed $b>0$
\begin{equation}
\log\psi_b(x)\sim 0,\quad\quad\quad (\text{as }
|x|\to\infty)\label{eq:psiAsy}
\end{equation}
with an exponentially small error, of the type $e^{-|x|}$---see
Corollary~2.3 of Rains \cite{Rains:2009} for the precise statement
and see Appendix~B of \cite{Ruijsenaars:1999} for an earlier
analysis in a different notation. This property of $\psi_b$
guarantees that the infinite product in (\ref{eq:GammaFVSq}) is
convergent when $\mathrm{Re}\beta>0$.

For $x$ strictly inside the strip
\begin{equation}
S^+: \quad
0<\mathrm{Re}(xe^{-i\mathrm{arg}\beta})<\mathrm{Re}(e^{-i\mathrm{arg}\beta}),\label{eq:S+def}
\end{equation}
as $|\beta|\to 0$ with $|\mathrm{arg}\beta|<\pi/2$ and $b>0$ fixed,
all the $\psi_b$ functions on the RHS of (\ref{eq:GammaFVSq})
approach unity exponentially fast. Moreover, the dominant piece of
$Q_+$ in the limit is of order $\frac{1}{\tau\sigma}$ and gives
\begin{equation}
    \Gamma(e^{2\pi i x})= e^{-2\pi i\frac{2x^3-3x^2+x}{12\tau\sigma}+\mathcal{O}(\frac{\tau+\sigma}{\tau\sigma})}\quad (\text{for $x\in S^+$}).\label{eq:leadingEstimate01}
\end{equation}
Since the LHS of the above relation is periodic in $x\to x+1$, we
can extend it beyond $x\in S^+$ by replacing every $x$ on the RHS
with its horizontal shift $\{x\}:=x-\lfloor
\mathrm{Re}x+\mathrm{Im}x\cdot\tan(\mathrm{arg}\beta)\rfloor$ to
inside $S^+$. For $x\in\mathbb{R}$ we have $\{x\}=x-\lfloor
x\rfloor$; this yields our desired estimate
(\ref{eq:leadingEstimate}).

Equivalently, we could use Proposition~2.11 of Rains
\cite{Rains:2009}, after identifying $v_{\mathrm{there}}$ with
$|\beta|/2\pi$, and $\omega_{1,2\ \mathrm{there}}$ with
$ib^{\pm 1}e^{i\mathrm{arg}\beta}$.\\

A somewhat subtle point is that the estimate
(\ref{eq:leadingEstimate}) is not uniform with respect to $x$ when
applied to the (``vector multiplet'') gamma functions in the
denominator of the RHS of (\ref{eq:EHI})---or more generally
(\ref{eq:leadingEstimate01}) is not uniform when $x$ approaches the
boundaries of the strip $S^+$. We need a uniform estimate because we
want to apply the estimate in the integrand of the index;
\emph{c.f.} the paragraph of Eq.~(3.15) in \cite{Ardehali:2015c}. We
expect though that an argument similar to that in the paragraph
below Eq.~(3.30) of \cite{Ardehali:2015c} can be given implying that
the non-uniform estimate introduces a negligible error on the
leading asymptotics of the index; we postpone the rigorous analysis
of this point to the future.

\subsection*{Cardy-like asymptotics of the index (\ref{eq:EHIasy1})}

It follows from the relation between the Pochhammer symbol and the
Dedekind eta function
\begin{equation}
\eta(\tau)=e^{2\pi i \tau/24}(e^{2\pi i \tau};e^{2\pi i \tau}),
\label{eq:etaPoc}
\end{equation}
and the modular property $\eta(-1/\tau)=\sqrt{-i\tau}\eta(\tau)$ of
the eta function that in the Cardy-like limit the Pochhammer symbols
on the RHS of (\ref{eq:EHI}) contribute an exponential growth of
\begin{equation*}
(p;p)^{N-1}(q;q)^{N-1}=
e^{\mathcal{O}(\frac{\tau+\sigma}{\tau\sigma})}.
\end{equation*}
They can hence be neglected, along with the $N!$ in the denominator
of (\ref{eq:EHI}), in the Cardy-like limit when
$0<|\mathrm{arg}\beta|<\pi/2$. We thus end up with
(\ref{eq:EHIsimplified}) as promised.\footnote{More precisely, we
also have to show that the complex phase of the integrand, arising
from the subleading terms that we have ignored, does not cause a
\emph{completely destructive interference} modifying the leading
asymptotics; \emph{c.f.} the paragraph below that of Eq.~(3.15) in
\cite{Ardehali:2015c}. For non-chiral SCFTs without flavor
fugacities, the absence of such cancelations was established in
\cite{Ardehali:thesis}. For the case at hand, we expect such
``unnatural'' completely destructive interferences to be absent at
least for generic $\mathrm{Re}\Delta_{1,2}$; we postpone the
rigorous analysis of this point to the future.}

We remind the reader that if $\beta\in\mathbb{R}_{>0}$ the integrand
of (\ref{eq:EHIsimplified}) becomes a pure phase, and the more
precise asymptotic analysis of Section~\ref{sec:betaReal} has to be
performed.\\

For $0<|\mathrm{arg}\beta|<\frac{\pi}{2}$, in the small-$|\beta|$
limit the integral $(\ref{eq:EHIsimplified})$ is localized around
the minima of $-\sin(2\mathrm{arg}\beta)\cdot
Q_h(\boldsymbol{x};\mathrm{Re}\Delta_k)$, whose
$\boldsymbol{x}$-dependent part can be read from (\ref{eq:QhDef}) to
be
\begin{equation}
\begin{split}
-\frac{\sin(2\mathrm{arg}\beta)}{12}\sum_{1\le i<j\le
N}\sum_{k=1}^3\kappa(\mathrm{Re}\Delta_k\pm
x_{ij}),\label{eq:QhXdep}
\end{split}
\end{equation}
with $x_{ij}=x_i-x_j$. The expression
\begin{equation}
\begin{split}
V^Q(x_{ij};\mathrm{arg}\beta,\mathrm{Re}\Delta_k)=-\sin(2\mathrm{arg}\beta)\cdot\sum_{k=1}^3\kappa(\mathrm{Re}\Delta_k\pm
x_{ij}),\label{eq:QhPot}
\end{split}
\end{equation}
is thus roughly a pair-wise potential for the ``holonomies'' $x_i$.

\begin{figure}[t]
\centering
    \includegraphics[scale=.5]{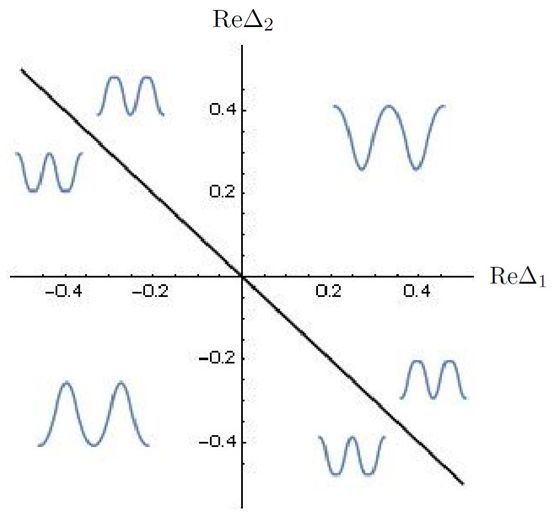}
\caption{The qualitative behavior of
$V^Q(x;\mathrm{arg}\beta,\mathrm{Re}\Delta_k)$ as a function of $x$
for fixed $\mathrm{Re}\Delta_{1,2}$ and fixed
$\mathrm{arg}\beta\in(-\pi/2,0)$, in various regions of the space of
the control-parameters $\mathrm{Re}\Delta_{1,2}$.
\label{fig:Catastrophe}}
\end{figure}

\begin{figure}[t]
\centering
    \includegraphics[scale=.5]{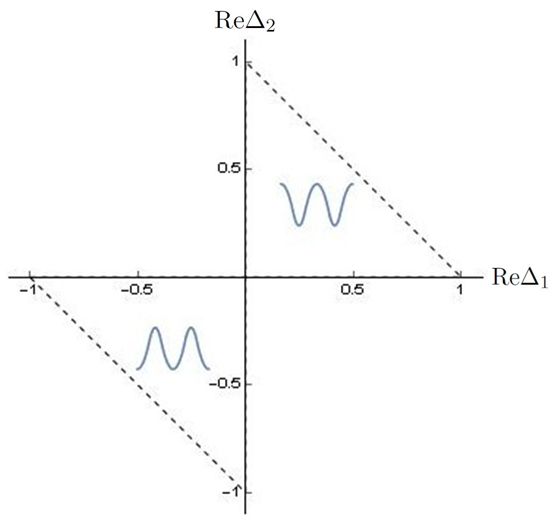}
\caption{The qualitative behavior of
$V^Q(x;\mathrm{arg}\beta,\mathrm{Re}\Delta_k)$, as a function of $x$
for fixed $\mathrm{Re}\Delta_{1,2}$ and fixed
$\mathrm{arg}\beta\in(-\pi/2,0)$, in the two complementary regions
$-1<\mathrm{Re}\Delta_{1},\mathrm{Re}\Delta_{2},-1-\mathrm{Re}\Delta_1-\mathrm{Re}\Delta_2<0$
(lower-left) and
$0<\mathrm{Re}\Delta_{1},\mathrm{Re}\Delta_{2},1-\mathrm{Re}\Delta_1-\mathrm{Re}\Delta_2<1$
(upper-right) of the space of the control-parameters
$\mathrm{Re}\Delta_{1,2}$. The $M$ and $W$ wings switch places if
$\mathrm{arg}\beta$ is taken to be inside $(0,\pi/2)$ instead.
\label{fig:CatastSimp}}
\end{figure}

We take $\mathrm{arg}\beta$ and $\mathrm{Re}\Delta_{1,2}$ to be our
control-parameters; $\mathrm{Re}\Delta_{3}$ is determined (mod
$\mathbb{Z}$ to be precise, which is enough) by the balancing
condition. We take the fundamental region of
$\mathrm{Re}\Delta_{1,2}$ to be $[-1/2,1/2]$. The two qualitatively
different behaviors that the function $V^Q$ can exhibit in various
regions of the space of the control-parameters
$\mathrm{Re}\Delta_{1,2}$ are shown in Figure~\ref{fig:Catastrophe}
for $-\pi/2<\mathrm{arg}\beta<0$. This figure can be deduced either
by numerically scanning (using Mathematica for instance) the
fundamental region $\mathrm{Re}\Delta_{1,2}\in[-1/2,1/2]$, for some
fixed $\mathrm{arg}\beta\in(-\pi/2,0)$, or by analytically
investigating the function
$\sum_{k=1}^3\kappa(\mathrm{Re}\Delta_k\pm x_{ij})$ in its various
regions of analyticity. Note that an $M$-type potential means
$x_{ij}=0$ is preferred in the small-$|\beta|$ limit, while a
$W$-type potential means some $x_{ij}\neq0$ (always a neighborhood
of $x_{ij}=\pm 1/2$ it turns out) is preferred. Since
Figure~\ref{fig:Catastrophe} is a bit too featureful, we use the
equivalence $\mathrm{Re}\Delta_{1,2}\to\mathrm{Re}\Delta_{1,2}\pm1$
to shift its triangular regions so that the equivalent
Figure~\ref{fig:CatastSimp} is obtained, which is one of the main
results of the present paper. It should be clear from the
$\sin(2\mathrm{arg}\beta)$ factor in (\ref{eq:QhPot}) that the $M$
and $W$ wings in Figure~\ref{fig:CatastSimp} switch places if
$\mathrm{arg}\beta$ is taken to be inside $(0,\pi/2)$ instead.

To be specific, let us continue with the
$\mathrm{arg}\beta\in(-\pi/2,0)$ case for the moment. Then on the
$M$ wing of Figure~\ref{fig:CatastSimp} the minimum value of $V^Q$
occurs at $x=0$. Moreover, since $V^Q$ is stationary at $x=0$, the
phase of the integral (\ref{eq:EHIsimplified}) is stationary there.
We conclude that the $x_{ij}=0$ configurations---which in our
SU($N$) case correspond to $x_j=\frac{n}{N}$ for some
$n\in\{0,1,\dots,N-1\}$ independent of $j$---indeed dominate the
leading small-$\beta$ asymptotics of the index for
$-\frac{\pi}{2}<\mathrm{arg}\beta<0$ and
$-1<\mathrm{Re}\Delta_{1},\mathrm{Re}\Delta_{2},-1-\mathrm{Re}\Delta_1-\mathrm{Re}\Delta_2<0$;
this proves that CKKN's conjecture in \cite{Choi:2018} is valid in
the range of parameters just mentioned, and fills the gap in their
derivation of the HHZ function in this region of the
parameter-space. (Note that in our SU($N$) case each choice of $n$
completely breaks the $\mathbb{Z}_N$ center symmetry generated by
$x_j\to x_j+\frac{1}{N}$, thus meeting deconfinement expectations.)

On the bifurcation set, indicated by the dashed lines in
Figure~\ref{fig:CatastSimp}, the functions $V^Q$ and $Q_h$ vanish; a
more precise analysis using the techniques of
Section~\ref{sec:betaReal} is then required, but in any case it is
clear that the asymptotic growth of the index is much slower (with
$\mathrm{Re}\log\mathcal{I}=\mathcal{O}(\frac{1}{|\beta|})$) there,
so we do not discuss this set any further.

The question we would like to address now is: do we have a faster or
a slower asymptotic growth on the $W$ wing? Here by scanning (using
Mathematica for instance) the whole range
$-1/2<\mathrm{Re}\Delta_{1,2}<1/2$ we realize that the minima
$\sum\kappa(+1/3\pm 1/2)=-5/9$ at $\mathrm{Re}\Delta_{1,2}=+1/3$ and
$x_{ij}=\pm1/2$ are lower than the minimum $\sum\kappa(-1/3\pm
0)=-4/9$ at $\mathrm{Re}\Delta_{1,2}=-1/3$ and $x_{ij}=0$. Does this
mean that the index exhibits a faster asymptotic growth at
$\mathrm{Re}\Delta_{1,2}=+1/3$? The answer turns out to be no for
$N=2$, and seems to be no for all $N>2$ as well.

For $N=2$ the reason is that the $\boldsymbol{x}$-independent piece
of $Q_h$ in (\ref{eq:QhDef}) moves $\sum\kappa(-1/3\pm 0)=-4/9$
further down by $-2/9$, while it moves $\sum\kappa(+1/3\pm
1/2)=-5/9$ further up by $+2/9$. Thus in the CKKN limit with
$-\pi/2<\mathrm{arg}\beta<0$ we have
\begin{equation}
    \begin{split}
        \mathcal{I}_{N=2}(p,q,y_{1,2,3})\xrightarrow{\text{$\mathrm{Re}\Delta_k=-\frac{1}{3}$}}
        e^{\frac{2\pi i}{12\tau\sigma}\cdot\frac{2}{3}},
    \end{split}\label{eq:N=2at-1/3}
\end{equation}
while
\begin{equation}
    \begin{split}
         \mathcal{I}_{N=2}(p,q,y_{1,2,3})\xrightarrow{\text{$\mathrm{Re}\Delta_k=+\frac{1}{3}$}}
        e^{\frac{2\pi i}{12\tau\sigma}\cdot\frac{1}{3}}.
    \end{split}\label{eq:N=2at+1/3}
\end{equation}
In short, for $N=2$ the fastest asymptotic growth in the CKKN limit
with $-\pi/2<\mathrm{arg}\beta<0$ occurs on the $M$ wing of
Figure~\ref{fig:CatastSimp}.

For higher ranks there is a more important reason why points on the
$W$ wing can not compete with the fastest asymptotic growth on the
$M$ wing. That is because for $N>2$ it is impossible to distribute
$N$ holonomies $x_i$ on the fundamental region $[-1/2,1/2]$ (with
$-1/2$ and $1/2$ identified of course, and with $x_N$ determined
from the rest via $\sum_{i=1}^N x_i\in\mathbb{Z}$) and have all of
them at equal distance $|x_{ij}|=1/2$ from each other. Colloquially
speaking, it is not possible to capitalize on the minima of $V^Q$ on
the $W$ wing at $|x_{ij}|=1/2$ with all the holonomies, whereas it
is possible to do so on the minima at $|x_{ij}|=0$ on the $M$ wing;
hence as we increase the rank it becomes more and more intuitively
likely that the fastest asymptotic growth of the index should occur
on the $M$ wing, and so we expect that only this region potentially
bears entropy functions of the AdS$_5$ blackholes.

Let us recapitulate our findings so far. We have demonstrated that
the $|x_{ij}|=0$ points are preferred in the CKKN limit on the $M$
wing of the space of the control-parameters
$\mathrm{Re}\Delta_{1,2}$, and thus the asymptotic result
(\ref{eq:EHIasy1}) is valid there. We have also argued intuitively
that the $W$ wing yields slower asymptotic growth and is not
expected to bear entropy functions as large as those of the AdS$_5$
blackholes.\\

It is straightforward to deduce the analogous statements for
$0<\mathrm{arg}\beta<\frac{\pi}{2}$. In that case the $M$ and $W$
wings of Figure~\ref{fig:CatastSimp} are swapped. Hence this time it
is on the upper-right wing that the $|x_{ij}|=0$ configurations are
preferred in the CKKN limit, and the asymptotic result
(\ref{eq:EHIasy1}) is valid, though this time with
$\mathrm{Re}\Delta_{3}=1-\mathrm{Re}\Delta_{1}-\mathrm{Re}\Delta_{2}$.
We also know that for $N=2$ the fastest asymptotic growth of the
index occurs on the upper-right wing, and we expect the same to be
true as $N$ increases.

\subsection*{The blackhole saddle-points (\ref{eq:CKKNHHZsaddle}),(\ref{eq:NEWsaddle})}

Now we ask: in the case $-\pi/2<\mathrm{arg}\beta<0$ does the
lower-left wing, and in the case $0<\mathrm{arg}\beta<\frac{\pi}{2}$
does the upper-right wing contain blackhole saddle-points? To make
contact with the AdS$_5$ blackholes we have to find the critical
points of the Legendre transform of $\log\mathcal{I}$ in the CKKN
limit. In both cases it turns out that one blackhole saddle-point
exists. The latter saddle-point seems to have been overlooked in
\cite{Choi:2018}, but can be obtained with minor modification of the
computations in their Section~2.3 as we now outline. Recall that
when $0<\mathrm{arg}\beta<\frac{\pi}{2}$ we impose $\sum_k
\Delta_k=\tau+\sigma+1$ rather than $\sum_k \Delta_k=\tau+\sigma-1$;
while CKKN \cite{Choi:2018} (following HHZ \cite{Hosseini:2017})
impose the latter relation via
\begin{equation}
\Delta_k=\frac{-z_k}{1+z_1+z_2+z_3+z_4},\quad
\sigma=\frac{z_4}{1+z_1+z_2+z_3+z_4},\quad
\tau=\frac{1}{1+z_1+z_2+z_3+z_4},\label{eq:constCKKN}
\end{equation}
the former relation can be simply imposed by putting
\begin{equation}
\Delta_k=\frac{z^\ast_k}{1+z^\ast_1+z^\ast_2+z^\ast_3+z^\ast_4},\quad
\sigma=\frac{-z^\ast_4}{1+z^\ast_1+z^\ast_2+z^\ast_3+z^\ast_4},\quad
\tau=\frac{-1}{1+z^\ast_1+z^\ast_2+z^\ast_3+z^\ast_4}.\label{eq:constUS}
\end{equation}
We now would like to argue that the $z^\ast_{1,2,3,4}$ which solve
the extremization problem for $0<\mathrm{arg}\beta<\frac{\pi}{2}$
are indeed the complex conjugates of the $z_{1,2,3,4}$ that CKKN
found solving the extremization problem for
$-\pi/2<\mathrm{arg}\beta<0$. To demonstrate this, we present some
of the details of the extremization problem, in parallel with
Section~2.3 of CKKN \cite{Choi:2018}. Setting the derivatives of
(\ref{eq:ext}) with respect to $z^\ast_{1,2,3,4}$ to zero, we get
\begin{equation}
Q_k+J_1=-\frac{N^2-1}{2}\frac{z^\ast_1z^\ast_2z^\ast_3}{z^\ast_4}\left(\frac{1}{z^\ast_k}+\frac{1}{z^\ast_4}\right),\quad
J_2-J_1=\frac{N^2-1}{2}\frac{z^\ast_1z^\ast_2z^\ast_3}{z^\ast_4}\left(\frac{1}{z^\ast_4}-1\right),\label{eq:CKKN2.75}
\end{equation}
similar to the CKKN case---c.f. their Eq.~(2.75). However,
extremization with respect to $z^\ast_4$ yields
\begin{equation}
S=2\pi
i(\frac{N^2-1}{2}\frac{z^\ast_1z^\ast_2z^\ast_3}{z^\ast_4}+J_2),
\end{equation}
with a different sign on the RHS compared to the CKKN case---c.f.
their Eq.~(2.79). As a result, the equations for $z^\ast_{1,2,3,4}$
following from the above relations read
\begin{equation}
z^\ast_{k}=-\frac{-S+2\pi i J_2}{-S-2\pi i Q_k},\quad
z^\ast_{4}=\frac{-S+2\pi i J_2}{-S+2\pi i J_1},
\end{equation}
with only an $S\to-S$ difference compared to the CKKN case---c.f.
their Eq.~(2.88). In particular, since $S$ is real, our $z^\ast$s
are complex conjugates of their $z$s, as claimed. The relations
(\ref{eq:CKKNHHZsaddle}) and (\ref{eq:NEWsaddle}) follow rather
effortlessly from (\ref{eq:constCKKN}) and (\ref{eq:constUS})
respectively.

To obtain $S$, we can follow CKKN and write things in terms of
$f^\ast:=\frac{N^2-1}{2}\frac{z^\ast_1z^\ast_2z^\ast_3}{z^\ast_4}$,
so that Eq.~(\ref{eq:CKKN2.75}) becomes
\begin{equation}
\frac{1}{z^\ast_{4}}=\frac{J_2-J_1}{f^\ast}+1,\quad
\frac{1}{z^\ast_{k}}=-\frac{Q_k+ J_2}{f^\ast}-1,
\end{equation}
and then use the definition of $f^\ast$ to obtain
\begin{equation}
f^\ast=-\frac{N^2-1}{2}\frac{{f^\ast}^2(J_2-J_1+f^\ast)}{(Q_1+J_2+f^\ast)(Q_2+J_2+f^\ast)(Q_3+J_2+f^\ast)},
\end{equation}
which is a cubic relation for $f^\ast$. The cubic equation that
follows for $S=2\pi i(f^\ast+J_2)$ will then have the entropy
functions (\ref{eq:entropy1}) and (\ref{eq:entropy2}) as its
solutions.\\

To demonstrate the self-consistency of our computations we need to
show that the CKKN/HHZ saddle-point (\ref{eq:CKKNHHZsaddle}) is
indeed on the lower-left wing of Figure~\ref{fig:CatastSimp} and has
$-\pi/2<\mathrm{arg}\beta<0$, while the new saddle-point lies on the
upper-right wing and has $0<\mathrm{arg}\beta<\frac{\pi}{2}$. We
show only the second statement, as the first follows using the fact
that the two saddle-points have their $\Delta_k,\tau,\sigma$
negative complex conjugate of each other.

A quick way to the desired result is to note that $S+2\pi i Q_k$ are
on a straight line in the complex plane, so that their reciprocals
are on a circle. This observation motivates the change of variables
$\frac{1}{S+2\pi i Q_k}=\frac{1}{2S}(1+e^{-i\phi_k})$, with
$\phi_k\in(0,\pi)$. Then the desired ranges of $\mathrm{Re}\Delta_k$
and $\mathrm{arg}\beta$ follow easily from (\ref{eq:NEWsaddle}) and
the vector representation of the complex numbers $1+e^{-i\phi_k}$,
after neglecting the subleading terms in the CKKN scaling limit
$J^2\sim Q^3\to\infty$.\\

In summary, we have shown that when
$0<\mathrm{arg}\beta<\frac{\pi}{2}$ a blackhole saddle-point exists
on the upper-right wing of Figure~\ref{fig:CatastSimp}; as
comparison of Eqs.~(\ref{eq:constCKKN}) and (\ref{eq:constUS})
shows, the new saddle-point has fugacities $p^\ast,q^\ast,y_k^\ast$
that are complex conjugates of the fugacities at the CKKN/HHZ
saddle-point. Moreover, we have argued that besides this and the
CKKN/HHZ saddle-point no other saddle-points with such large
entropies exist in the Cardy-like limit.

\subsection*{Moving the flavor fugacities away from the unit circle}

As we noted at the end of Subsection~\ref{subsec:outline}, unless
$Q_1=Q_2=Q_3$, the critical $\Delta_k$ have nonzero imaginary parts,
and thus the critical fugacities $u_k$ (and also $y_k$) lie away
from the unit circle. Hence to complete the blackhole entropy
derivation for the general case with unequal $Q_k$, we need to be
able to justify the Cardy-like asymptotics (\ref{eq:EHIasyHHZ}) when
$\mathrm{Im}\Delta_k$ are not sent to zero.

A partial justification is as follows. Let us assume that
$\mathrm{Im}\Delta_k$ are small enough so that the integral
(\ref{eq:EHI}) still represents the index, albeit with a slightly
deformed contour of integration\footnote{It appears like we might
only need the contour-deformation to be small near
$x_j=\frac{n}{N}$, which are the dominant eigenvalue configurations
in the regime of parameters pertaining to the blackhole
saddle-points.}. We can then use (\ref{eq:leadingEstimate01}) to
arrive at the following variant of (\ref{eq:EHIsimplified}):
\begin{equation}
    \begin{split}
        \mathcal{I}(p,q,y_{1,2,3})\xrightarrow{\ }\int_{-1/2}^{1/2}\mathrm{d}^{N-1}\boldsymbol{x}\
        e^{-2\pi i\frac{Q_h(\boldsymbol{x};\mathrm{arg}\beta,\Delta_k)}{\tau\sigma}},
    \end{split}\label{eq:EHIsimplifiedGen}
\end{equation}
with
\begin{equation}
\begin{split}
Q_h(\boldsymbol{x};\mathrm{arg}\beta,\Delta_k):=\frac{1}{12}\sum_{k=1}^3\left((N-1)\kappa(\Delta_k)+\sum_{1\le
i<j\le N}\kappa(\Delta_k\pm (x_i-x_j))\right),\label{eq:QhDefGen}
\end{split}
\end{equation}
where $\kappa(x)$ is still defined as in (\ref{eq:kappaDef}), but
with $\{x\}:=x-\lfloor
\mathrm{Re}x+\mathrm{Im}x\cdot\tan(\mathrm{arg}\beta)\rfloor$ as
discussed around (\ref{eq:leadingEstimate01}).

We expect that for fixed $\mathrm{arg}\beta$ (either in $(-\pi/2,0)$
or in $(0,\pi/2)$), and for small enough $\mathrm{Im}\Delta_k$, the
catastrophic behavior of the pair-wise potential for the holonomies
to remain similar to that discussed above, with the two
complementary ``wings'' $\Delta_{1,2},1-\Delta_1-\Delta_2\in S^+$
and $\Delta_{1,2},-1-\Delta_1-\Delta_2\in S^+ -1$ being associated
to $M$- or $W$-type behaviors, with one or the other having $x=0$ as
its preferred configuration depending on the sign of
$\mathrm{arg}\beta$. Then for $\mathrm{arg}\beta\in(0,\pi/2)$ one
can use (\ref{eq:leadingEstimate01}) on the wing
$\Delta_{1,2},1-\Delta_1-\Delta_2\in S^+$ to arrive at
(\ref{eq:EHIasyHHZ}) with $\sum_k \Delta_k=\tau+\sigma+1$, while for
$\mathrm{arg}\beta\in(-\pi/2,0)$ one can use
(\ref{eq:leadingEstimate01}) with $x\to x+1$ on the wing
$\Delta_{1,2},-1-\Delta_1-\Delta_2\in S^+ -1$ to arrive at
(\ref{eq:EHIasyHHZ}) this time with $\sum_k \Delta_k=\tau+\sigma-1$.

Beyond a small neighborhood of $\mathrm{Im}\Delta_k=0$ the methods
of the present paper do not seem powerful enough to demonstrate
(\ref{eq:EHIasyHHZ}). Whether the fascinating formalism of
\cite{Closset:2017Bethe,Benini:2018Bethe} can help addressing the
general case with nonzero $\mathrm{Im}\Delta_k$ is currently being
investigated.

%
%
%
\section{Real-valued temperature ($\beta\to0$, $\beta\in\mathbb{R}_{>0}$)}\label{sec:betaReal}

In this section we keep the spacetime fugacities $p,q$ real-valued
and define $b,\beta\in\mathbb{R}_{>0}$ through $p=e^{-\beta
b},q=e^{-\beta b^{-1}}$. We also keep the flavor fugacities
$u_k=e^{2\pi i T_k}$ on the unit circle (hence $T_k\in\mathbb{R}$),
and study the effect of finite nonzero $T_k$ on the small-$\beta$
asymptotics of the index.

In order to provide some conceptual context for the somewhat
technical analysis in the rest of this section we now briefly
discuss the path-integral interpretation of the index with
real-valued $p,q$. We will still be analyzing the Hamiltonian index
$\mathcal{I}$, and only importing intuition from the path-integral
picture---until the next section where the path-integral partition
function is analyzed.

The superconformal index with real $p,q$ can be obtained via the
path-integral SUSY partition function of the theory on $S_b^3\times
S^1_\beta$, where $S_b^3$ is the squashed three-sphere with unit
radius and squashing parameter $b$, while $S^1_\beta$ is the circle
with circumference $\beta$ \cite{Assel:2014}. The integration
variables $z_i$ in the index (\ref{eq:EHI}) correspond to the
eigenvalues of the holonomy matrix $P\exp(i\oint_{S^1_\beta}A_0)$,
with $A_0$ the component along $S^1_\beta$ of the SU($N$) gauge
field. The $u_k$ correspond to the eigenvalues of the background
holonomy matrix $P\exp(i\oint_{S^1_\beta}A^u_0)$, with $A^u_0$ the
component along $S^1_\beta$ of the background gauge field $A^u$
associated to the ``flavor'' SU($3$) of the $\mathcal{N}=4$ theory.
The path-integral partition function is actually a Casimir-energy
factor different from the index; this factor is irrelevant for the
present analysis and we postpone its discussion to the next section.
Interpreting the $S_b^3$ as the spatial manifold and the $S^1_\beta$
as the Euclidean time circle, we refer to $\beta$ as the
inverse-temperature in analogy with thermal quantum physics---even
though our fermions have supersymmetric (\emph{i.e.} periodic)
boundary conditions around $S^1_\beta$.

Next, we note that while large-$N$ QFTs ($N\to\infty$) on compact
spatial manifolds can have finite-temperature phases associated to
large-$N$ saddle-points, in the present work we are considering a
finite-$N$ QFT on a compact spatial manifold (namely $S_b^3$), which
can not be assigned a phase at any finite temperature. In the
high-temperature limit ($\beta\to0$), however, infinite-temperature
phases can be associated to the small-$\beta$ saddle-points. In
particular, we will say that the infinite-temperature phase of the
index is \emph{Higgsed} if the dominant small-$\beta$
saddle-point(s) of its matrix-integral lie away from the ``origin''
$\boldsymbol{x}=0$. For example, the infinite-temperature phase of
the index of the SU($2$) ISS model is Higgsed, but that of the
$\mathcal{N}=1$ SU($N$) SQCD (say in the conformal window) is not
\cite{Ardehali:2015c}.

Moreover, we will say that the infinite-temperature phase of the
index is \emph{exponentially growing} if for the leading
small-$\beta$ asymptotics we have $\mathrm{Re}\log\mathcal{I}\approx
A/\beta$ with $A>0$; in other words if the index exhibits
exponential \emph{growth} in the high-temperature limit.

Below we will see that for generic non-zero $T_k\in\mathbb{R}$ the
infinite-temperature phase of the index of the SU($N$)
$\mathcal{N}=4$ theory is Higgsed, and in the $N=2$ case for some
specific range of $T_k$ also exponentially growing.

Taking $p,q$ to be real means taking $\tau,\sigma$ to be pure
imaginary. Then we have $\mathrm{Im}(\tau\sigma)=0$, so that the
estimate (\ref{eq:leadingEstimate}) gives only a pure phase; we thus
have to consider the subleading terms in the exponent of its RHS to
get information about the modulus of the index. The improved
estimate is \cite{Ardehali:2015c}
\begin{equation}
\begin{split}
\log\Gamma\big((pq)^{r/2}e^{2\pi i x}\big)= 2\pi
i(-\frac{\kappa(x)}{12\tau\sigma}
+(r-1)\frac{\tau+\sigma}{4\tau\sigma}\vartheta(x)-(r-1)\frac{\tau+\sigma}{24\tau\sigma})+O(\tau^0,\sigma^0)&\\
(\text{for $r\in(0,2)$\ and
$x\in\mathbb{R}$})&,\label{eq:GammaOffCenter2}
\end{split}
\end{equation}
where the continuous, positive, even, periodic function
\begin{equation}
\begin{split}
\vartheta(x)&:=\{x\}(1-\{x\})\\
&\left(=|x|-x^2 \quad\quad\text{for
$x\in[-1,1]$}\right),\label{eq:varthetaDef}
\end{split}
\end{equation}
is defined after Rains \cite{Rains:2009}.

In order to apply the estimate (\ref{eq:GammaOffCenter2}) to the
gamma functions in (\ref{eq:EHI}) we have to interpret the modulus
of the arguments of the gamma functions as $(pq)^{r/2}$, and
interpret the phase of the arguments as $2\pi x$; then, for
instance, we can apply (\ref{eq:GammaOffCenter2}) to the gamma
function in the numerator of the integrand of (\ref{eq:EHI}) by
identifying $r$, $x$ as $r=2/3$, $x=T_k\pm(x_i-x_j)$; note that the
balancing condition $\prod_{k=1}^3 y_k=pq$ implies $\sum_{k=1}^3
T_k\in\mathbb{Z}$. Since the Pochhammer symbols in (\ref{eq:EHI})
yield asymptotics that cancel the contribution of the gamma
functions from the third term on the RHS of
(\ref{eq:GammaOffCenter2}) \cite{DiPietro:2014,Ardehali:2015c},
applying (\ref{eq:GammaOffCenter2}) to (\ref{eq:EHI}) we get
\begin{equation}
    \begin{split}
        \mathcal{I}(p,q,u_{1,2,3})\xrightarrow{\text{$p,q$ real, $|u_k|=1$}}\int_{-1/2}^{1/2}\mathrm{d}^{N-1}\boldsymbol{x}\
        e^{-\frac{(2\pi)^2}{\beta}\left(\frac{b+b^{-1}}{2}\right)L_h(\boldsymbol{x},r_k=2/3;T_k)+i\frac{(2\pi)^3}{\beta^2}Q_h(\boldsymbol{x};T_k)},
    \end{split}\label{eq:EHIsimplifiedRealT}
\end{equation}
where we have used $\tau=i\beta b^{-1}/2\pi$ and $\sigma=i\beta
b/2\pi$. The functions $L_h$ and $Q_h$ are the natural
generalizations of those defined in \cite{Ardehali:2015c} for
$T_k=0$, and are explicitly given by\footnote{Because of the ABJ
U(1)$_R$ anomaly cancelation $L_h$ is a piecewise ``L''inear
function of $\boldsymbol{x}$ (\emph{c.f.} \cite{Ardehali:2015c}).
See Figure~\ref{fig:genRainsSU2}.}
\begin{equation}
\begin{split}
L_h(\boldsymbol{x},r_k=2/3;T_k)&=(N-1)\cdot\frac{1}{6}\big(\vartheta(T_1)+\vartheta(T_2)+\vartheta(T_3)\big)\\
&\ \ \ \ +\sum_{1\le i<j\le
N}\frac{1}{6}[\vartheta(x_i-x_j+T_1)+\vartheta(x_i-x_j-T_1)\\
&\ \ \ \ \ \ \ \ \ \ \ \ \ \ \ \ \ \ \ \
+\vartheta(x_i-x_j+T_2)+\vartheta(x_i-x_j-T_2)\\
&\ \ \ \ \ \ \ \ \ \ \ \ \ \ \ \ \ \ \ \
+\vartheta(x_i-x_j+T_3)+\vartheta(x_i-x_j-T_3)]-\vartheta(x_i-x_j).
\end{split}\label{eq:genRainsN}
\end{equation}
\begin{equation}
\begin{split}
Q_h(\boldsymbol{x};T_k)&=(N-1)\cdot\frac{1}{12}\big(\kappa(T_1)+\kappa(T_2)+\kappa(T_3)\big)\\
&\ \ \ \ +\sum_{1\le i<j\le
N}\frac{1}{12}\big(\kappa(x_i-x_j+T_1)-\kappa(x_i-x_j-T_1)\\
&\ \ \ \ \ \ \ \ \ \ \ \ \ \ \ \ \ \ \ \
+\kappa(x_i-x_j+T_2)-\kappa(x_i-x_j-T_2)\\
&\ \ \ \ \ \ \ \ \ \ \ \ \ \ \ \ \ \ \ \
+\kappa(x_i-x_j+T_3)-\kappa(x_i-x_j-T_3)\big).
\end{split}\label{eq:genQhN}
\end{equation}
Note that for $T_k=0$ both functions identically vanish, as in
\cite{Ardehali:2015c}.

Since the $1/\beta^2$ term in the exponent of the RHS of
(\ref{eq:EHIsimplifiedRealT}) gives a pure phase, the dominant
contribution to the integral presumably comes from the locus of
minima of $L_h(\boldsymbol{x},r_k=2/3;T_k)$. One has to make sure
that $Q_h(\boldsymbol{x};T_k)$ is stationary at that locus though,
otherwise a more careful analysis is required.

\subsection*{The SU$(2)$ case}
\begin{figure}[t]
\centering
    \includegraphics[scale=.47]{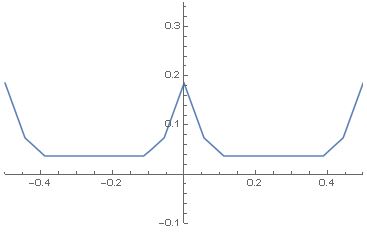}
    \hspace{.35cm}
    \includegraphics[scale=.47]{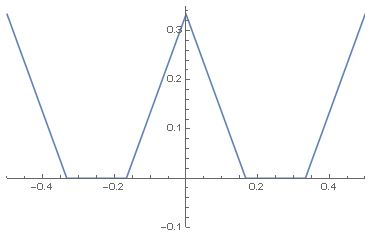}
    \hspace{.35cm}
    \includegraphics[scale=.47]{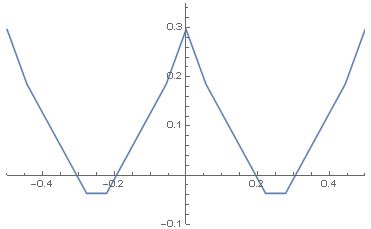}
\caption{The $L_h$ function (\ref{eq:genRainsN}) in the $N=2$ case,
as a function of $x_1$, for $T_{1,2}=-1/9$ (left), $T_{1,2}=-1/3$
(middle), and $T_{1,2}=-4/9$ (right). \label{fig:genRainsSU2}}
\end{figure}

Take for example the $N=2$ case. Figure~\ref{fig:genRainsSU2} shows
the $L_h$ function of the SU($2$) $\mathcal{N}=4$ theory for sample
values of $T_k$. As the picture clearly shows, at the point $x_1=0$
the integrand is maximally suppressed.

\begin{figure}[t]
\centering
    \includegraphics[scale=.8]{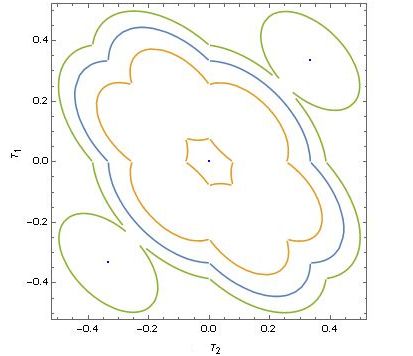}
\caption{Contours of $L_h(x_1=\pm1/4,r_k=2/3;T_k)$. The blue curve
and dots correspond to zero value, inside the blue curve except at
the origin corresponds to positive values, while outside the blue
curve and away from the blue dots corresponds to negative values.
\label{fig:Contours}}
\end{figure}

It is easy to check that the dominant configuration for $N=2$ is
$|x_1|=1/4$ (Figure~\ref{fig:genRainsSU2} is suggestive of this
also); not only $L_h$ is minimized there, but also $Q_h$ is
stationary as desired. Moreover, we see from
Figure~\ref{fig:genRainsSU2} that depending on $T_k$ the minimum of
$L_h$ can be positive, negative, or zero. Only when the minimum is
negative the infinite-temperature phase is exponentially growing.
The contours of $L_h(x_1=\pm1/4,r_k=2/3;T_k)$ are shown in
Figure~\ref{fig:Contours}: outside the blue contour we have
$L_h(x_1=\pm1/4,r_k=2/3;T_k)<0$, so the index is exponentially
growing, except on the blue dots at $T_1=T_2=\pm1/3$ where
$L_h(x_1=\pm1/4,r_k=2/3;T_k)$ vanishes.

Let us review what we have observed. While for $T_k=0$ both
functions $L_h$ and $Q_h$ are zero and the index has a power-law
asymptotics (more precisely an $\mathcal{I}\approx 1/\beta$ behavior
as $\beta\to 0$ \cite{Ardehali:2015c}), finite nonzero $T_k$ can
induce Mexican-hat potentials for the holonomies in the
high-temperature limit, triggering an infinite-temperature
exponential growth in the index.

\subsection*{Higher ranks}

We now show that the integrand of the index is maximally suppressed
at $\boldsymbol{x}=0$ in fact for arbitrary $N\ge2$ and
$T_k\in\mathbb{R}/\mathbb{Z}$.

Let us study the behavior of the $L_h$ function in
(\ref{eq:genRainsN}) with respect to $x_i$. For this purpose, we use
the following equality derived in \cite{Ardehali:2015c} (c.f.
Eq.~(3.51) there), valid for $-1/2\le u_i\le 1/2$:
\begin{equation}
\begin{split}
(2M-2)\sum_{1\le l\le M}\vartheta(u_l)&-\sum_{1\le l<m\le
M}\vartheta(u_l+u_m)-\sum_{1\le l<m\le
M}\vartheta(u_l-u_m)=2\sum_{1\le l<m\le
M}\mathrm{min}(|u_l|,|u_m|).\label{eq:SONvarthetaId}
\end{split}
\end{equation}
We will use the above identity with $M=4$ and $u_{1,2,3}=T_{1,2,3}$;
we would moreover like to take $u_4=x_i-x_j$, but this is not
allowed since the range $-1<x_i-x_j<1$ is incompatible with $-1/2\le
u_4\le 1/2$; to fix that we put instead $u_4=\{x_i-x_j+1/2\}-1/2$.
Using (\ref{eq:SONvarthetaId}) we can now rewrite the $L_h$ function
in (\ref{eq:genRainsN}) such that its only
$\boldsymbol{x}$-dependent piece is
\begin{equation}
-\frac{1}{3}\sum_{1\le i<j\le N}\sum_{k=1}^3
\mathrm{min}\big(|T_k|,|\{x_i-x_j+1/2\}-1/2|\big)
\end{equation}
The above expression is obviously negative-semi-definite as a
function of $x_i$, and it is maximized when $x_i-x_j=0$. So the
index is Higgsed for any $T_k\in\mathbb{R}/\mathbb{Z}$ at infinite
temperature.

Whether (or for which range of $T_k\in\mathbb{R}$) the
infinite-temperature phase of the index can be exponentially growing
when $N>2$, is an interesting problem which seems to require an
intricate analysis.

%
%
%

\section{Supersymmetric Casimir energy with complex chemical
potentials}\label{sec:Casimir}

When all the fugacities $p,q,u_k$ are real-valued, the index
$\mathcal{I}(p,q,u_k)$ is related to the path-integral SUSY
partition function $Z(\beta,b,m_k)$ of the theory on $S_b^3\times
S^1_\beta$ via \cite{Bobev:2015}
\begin{equation}
Z(\beta,b,m_k)=e^{-\beta
E_{\mathrm{SUSY}}(b,m_k)}\mathcal{I}(p,q,u_k),\label{eq:BBK}
\end{equation}
where $E_{\mathrm{SUSY}}(b,m_k)$ is known as the supersymmetric
Casimir energy, $\beta,b,m_k$ are defined through
\begin{equation}
p=e^{-\beta b},\quad q=e^{-\beta b^{-1}},\quad u_k=e^{-\beta m_k},
\end{equation}
and $S_b^3$ is the squashed three-sphere with unit radius and
squashing parameter $b$, while $S^1_\beta$ is the circle with
circumference $\beta$. (The special case of (\ref{eq:BBK}) with
$m_k=0$ was understood already in \cite{Ardehali:2015b,Assel:2015s},
based on earlier slightly contrasting computations of
\cite{Assel:2014}.)

As made clear by HHZ \cite{Hosseini:2017} (and further elucidated in
\cite{Choi:2018,Choi:2018b,Benini:2018,Cabo-Bizet:2018}) making
contact with the AdS$_5$ BPS blackholes requires considering complex
fugacities $p,q,u_k$ in the index. With the goal of understanding
the role of the supersymmetric Casimir energy in the blackhole
entropy discussion, in this section we study the relation between
$Z$ and $\mathcal{I}$ for complex fugacities such that
$b\in\mathbb{R}_{>0}$ and $\beta\in\mathbb{C}$ with
$\mathrm{Re}\beta>0$ as in Section~\ref{sec:betaComplex}, while
$u_k$ are on the unit circle as in Section~\ref{sec:betaReal}.
Rather than modifying the background geometry to achieve such
complexified $\beta$ (\emph{c.f.} \cite{Assel:2014}), we simply
analytically continue the results obtained for real $p,q$.

Let us consider a free chiral multiplet to begin with; as in
\cite{Assel:2014,Ardehali:2015b}, we expect that solving this case
leads to the solution of the interacting non-abelian case as well.

Following Appendix~A of \cite{Ardehali:2015b}, we start with the
one-loop determinant of the $n$th KK mode on $S^3\times S^1$.
Eq.~(A.15) in \cite{Ardehali:2015b} now generalizes to
\begin{equation}
\log Z^{(n)}=\ell_b\big(-(R-1)\frac{b+b^{-1}}{2}+\frac{2\pi
i}{\beta}(n+T_k)\big),
\end{equation}
where $\ell_b$ is the special function discussed in
\cite{Ardehali:2015b}, the $R$-charge of the multiplet is denoted by
$R$, and $T_k:=\frac{i\beta m_k}{2\pi}\in\mathbb{R}$, with $m_k$ the
only chemical potential the chiral multiplet couples to.

Define $X:=(R-1)\frac{b+b^{-1}}{2}$ for notational convenience.
Following \cite{Ardehali:2015b} step by step, we now rewrite $\log
Z^{(n)}$ in terms of $\psi_b$ which has a simple asymptotic
behavior. Eq.~(A.2) of \cite{Ardehali:2015b} implies that in terms
of $\psi_b$:
\begin{equation}
\log Z^{(n)}=\log \psi_b\big(X-\frac{2\pi
i}{\beta}(n+T_k)\big)+\frac{i\pi}{2}\left(X-\frac{2\pi
i}{\beta}(n+T_k)\right)^2-i\pi(\frac{b^2+b^{-2}}{24}).
\end{equation}
Now, using the fact that $\ell_b(-x)=-\ell_b(x)$, we can rewrite
\begin{equation}
\begin{split}
\log Z^{(n)}&=\log \psi_b\big([X-\frac{2\pi
i}{\beta}(n+T_k)]\mathrm{sgn}(n+T_k)\big)^{\mathrm{sgn}(n+T_k)}-\frac{i\pi}{2}\frac{4\pi^2}{\beta^2}\mathrm{sgn}(n+T_k)\big(n+T_k\big)^2\\
&\ \ \ \
+\frac{4\pi^2}{2\beta}\mathrm{sgn}(n+T_k)\big(n+T_k\big)X+[\frac{i\pi}{2}X^2-i\pi(\frac{b^2+b^{-2}}{24})]\mathrm{sgn}(n+T_k).\label{eq:nthLevel}
\end{split}
\end{equation}
One way to check the above equation is to check it separately for
$\mathrm{sgn}(n+T_k)=+1$ and $\mathrm{sgn}(n+T_k)=-1$, using
$\ell_b(-x)=-\ell_b(x)$ and Eq.~(A.2) in \cite{Ardehali:2015b}. The
reason for this rewriting is to divide $\psi_b$s into the numerator
and denominator of $Z$, so we can eventually relate $Z$ to
$\mathcal{I}$ using expressions such as (\ref{eq:GammaFVSq}).

Finally, we sum (\ref{eq:nthLevel})  over $n\in\mathbb{Z}$. In doing
so, we use the relations Di~Pietro and Honda used
\cite{DiPietro:2017} for analyzing the high-temperature asymptotics
of the index:
\begin{eqnarray}
\sum_{n\in\mathbb{Z}}\mathrm{sgn}(n+T_k)&=1-2\{T_k\},\label{eq:sum0}\\
\sum_{n\in\mathbb{Z}}\mathrm{sgn}(n+T_k)\big(n+T_k\big)&=\vartheta(T_k)-\frac{1}{6},\label{eq:sum1}\\
\sum_{n\in\mathbb{Z}}\mathrm{sgn}(n+T_k)\big(n+T_k\big)^2&=-\frac{1}{3}\kappa(T_k).\label{eq:sum2}
\end{eqnarray}
The definitions are $\{x\}:=x-\lfloor x\rfloor$,
$\vartheta(x):=\{x\}(1-\{x\})$,
$\kappa(x):=\{x\}(1-\{x\})(1-2\{x\})$.

With this regularization---combining techniques from
\cite{Ardehali:2015b} and \cite{DiPietro:2017}---we obtain
\begin{equation}
\begin{split}
\log Z&=\sum_{n\in\mathbb{Z}}\log \psi_b\big([X-\frac{2\pi
i}{\beta}(n+T_k)]\mathrm{sgn}(n+T_k)\big)^{\mathrm{sgn}(n+T_k)}\\
&\ \ \ \ +\frac{i(2\pi)^3}{12\beta^2}\kappa(T_k)+\frac{(2\pi)^2}{2\beta}(R-1)\frac{b+b^{-1}}{2}(\vartheta(T_k)-\frac{1}{6})\\
&\ \ \ \
+[\frac{i\pi}{2}\big((R-1)\frac{b+b^{-1}}{2}\big)^2-i\pi(\frac{b^2+b^{-2}}{24})](1-2\{T_k\}).\label{eq:summed}
\end{split}
\end{equation}
Putting $T_k=0$ we can compare with Eq.~(A.16) in
\cite{Ardehali:2015b}, noting that $\kappa(0)=\vartheta(0)=0$, so
that the only surviving term on the second line of the RHS of the
above relation gives the Di~Pietro-Komargodski asymptotics
\cite{DiPietro:2014} as $\beta\to0$; the first and the third lines
combine to give the first and the third terms on the RHS of
Eq.~(A.16) in \cite{Ardehali:2015b}.\\

We are done with our regularization. We believe our method of
regularization is correct because we have been careful with the
convergence of the infinite product appearing in $Z$---or
equivalently the convergence of the infinite sum appearing in $\log
Z$---after regularization, and because we have used well-established
tools of analytic continuation\footnote{See Chapter~VII of
\cite{Weil:1975} for some context.} for evaluating the sums
(\ref{eq:sum0})--(\ref{eq:sum2}). As a byproduct, from the second
line on the RHS of (\ref{eq:summed}) we can read off the
high-temperature asymptotics of the partition function of a chiral
multiplet with a flavor fugacity on the unit circle.

We now would like to relate $Z$ as obtained in (\ref{eq:summed}) to
the index $\mathcal{I}$. We use (\ref{eq:GammaFVSq}) and the fact
that the index of the chiral multiplet is $\Gamma((pq)^{R/2}u_k)$.
For simplicity we assume $0<T_k<1$, and replace all $\{T_k\}$ in
(\ref{eq:summed}) with $T_k$. Then we set (\ref{eq:summed}) equal to
\begin{equation}
\log\mathcal{I}-\beta E_{\mathrm{SUSY}}(b,T_k)=2\pi i
Q_+(x=R(\tau+\sigma)/2+T_k;\sigma,\tau)+\sum\log\psi_b-\beta
E_{\mathrm{SUSY}}(b,T_k).
\end{equation}
The end result is that $E_{\mathrm{SUSY}}$ comes out just as in
\cite{Ardehali:2015b,Assel:2015s}: there is no dependence on $T_k$!

In other words, for $u_k=e^{2\pi i T_k}$ on the unit circle (which
is relevant to the equal-charge AdS$_5$ blackholes) we have
$E_{\mathrm{SUSY}}(b,m_k)=E_{\mathrm{SUSY}}(b,0)$. Since in the
small-$|\beta|$ limit with $b>0$ fixed we have $\beta
E_{\mathrm{SUSY}}(b,0)\to0$, we conclude that on the saddle-point
associated to the equal-charge blackholes the supersymmetric Casimir
energy has no significance in the leading Cardy-like asymptotics of
the partition function $Z$. In particular, the Casimir-energy factor
relating $Z$ and $\mathcal{I}$ is irrelevant to the blackhole
entropy function arising in the Cardy-like limit of either.

The relation between the above discussion and the interesting
proposal of \cite{Cabo-Bizet:2018} which seems to involve analytic
continuation of $Z$ with respect to $\tau$ and $\sigma$ is currently
under study.

%
%
%
\section{Summary and open problems}\label{sec:open}

We have presented a careful analysis of the asymptotics of the
SU($N$) $\mathcal{N}=4$ theory index (\ref{eq:EHI}) in the CKKN
limit where the flavor fugacities approach the unit circle and the
spacetime fugacities approach $1$. We emphasize that compared to the
previous work \cite{Ardehali:2015c} the Cardy-like limit studied
here is more general in two respects: $i)$ instead of sending the
flavor fugacities to $1$ as in \cite{Ardehali:2015c}, following CKKN
here we allowed the flavor fugacities to approach the unit circle;
$ii)$ although in Section~\ref{sec:betaReal} we kept the
``inverse-temperature'' $\beta$ on the positive real axis as in
\cite{Ardehali:2015c}, in Section~\ref{sec:betaComplex} we
complexified $\beta$ and let $|\mathrm{arg}\beta|\in(0,\pi/2)$ which
was necessary for making contact with the HHZ function and the
AdS$_5$ blackholes.

For complexified temperature (with $0<|\mathrm{arg}\beta|<\pi/2$),
we have demonstrated that in the CKKN limit, the dominant holonomy
configurations in the index are dictated by the pairwise potential
(\ref{eq:QhPot}). We explained that depending on the sign of
$\mathrm{arg}\beta$, the pair-wise potential has $M$- or $W$-type
behavior on complementary wings of the space of the
control-parameters $\mathrm{Re}\Delta_{1,2}$---see
Figure~\ref{fig:CatastSimp}. On the $M$ wings the potential is
minimized at the origin, so the holonomies condense (at either of
$N$ possibilities $x_j=0,\frac{1}{N},\dots,\frac{N-1}{N}$ breaking
the $\mathbb{Z}_N$ center), thereby giving rise to the HHZ function
(\ref{eq:EHIasyHHZ}) (with
$\Delta_3=\tau+\sigma-\Delta_1-\Delta_2+\mathrm{sgn}(\mathrm{arg}\beta)$)
in the leading asymptotics of the index, from which we extracted two
blackhole saddle-points (one for each sign of $\mathrm{arg}\beta$).
On the other hand, on the $W$ wings, except for the $N=2$ case, the
quantitative analysis seems difficult; we only presented intuitive
arguments suggesting that for $N>2$ the index has a slower
asymptotic growth there and therefore no blackhole saddle-points
with entropies as large as the two just mentioned are expected in
those regions.

\begin{description}
  \item[Problem 1)] In the CKKN limit
\begin{equation}
|\sigma|,|\tau|,\mathrm{Im}\Delta_k\to 0,\ \text{with\
}\frac{\tau}{\sigma}\in\mathbb{R}_{>0},\mathrm{Re}\Delta_{k}\
\text{fixed},\ \text{and}\ \mathrm{Im}\tau,\mathrm{Im}\sigma>0,
\end{equation}
find the asymptotics of the SU($N$) $\mathcal{N}=4$ theory index for
$N>2$ on the $W$ wings; that is, for $\tau,\sigma$ inside the 2nd
quadrant and $\mathrm{Re}\Delta_{1,2}$ on the lower-left wing of
Figure~\ref{fig:CatastSimp}, or for $\tau,\sigma$ inside the 1st
quadrant and $\mathrm{Re}\Delta_{1,2}$ on the upper-right wing of
Figure~\ref{fig:CatastSimp}. In particular, show that the fastest
asymptotic growth in those regions is slower than the fastest growth
on the complementary $M$ wings.
\end{description}

Even without addressing the above problem, we have successfully
derived two blackhole saddle-points in the CKKN limit. However, the
saddle-points have flavor fugacities that are away from the unit
circle unless the three charges $Q_k$ are (approximately) equal [in
which case the critical $\Delta_{1,2,3}$ are in fact (approximately)
simply $\mathrm{sgn}(\mathrm{arg}\beta)\times \frac{1}{3}$].
Therefore our derivation of the blackhole entropy function is
incomplete for the general blackholes with unequal $Q_k$. To
complete the analysis for the general case we have to derive the
asymptotic relation (\ref{eq:EHIasyHHZ}) when $\mathrm{Im}\Delta_k$
are not sent to zero.

\begin{description}
  \item[Problem 2)] For complexified temperature ($0<|\mathrm{arg}\beta|<\pi/2$) perform the asymptotic analysis of the index in the limit
\begin{equation}
|\sigma|,|\tau|\to 0,\ \text{with\
}\frac{\tau}{\sigma}\in\mathbb{R}_{>0},\Delta_{k}\in\mathbb{C}\
\text{fixed},\ \text{and}\ \mathrm{Im}\tau,\mathrm{Im}\sigma>0.
\end{equation}
In particular, derive the HHZ function, once with $\sum_k
\Delta_k=\tau+\sigma-1$ and once with $\sum_k
\Delta_k=\tau+\sigma+1$, in two separate regimes of parameters.
\end{description}

We would like to emphasize that although we have not given a
complete derivation of the entropy function for the general case
with unequal charges, our analysis in the equal-charge case already
allows addressing various conceptual issues in the derivation. One
such conceptual issue has been the significance of the rather
special relation $\sum_k \Delta_k=\tau+\sigma-1$ in the HHZ function
\cite{Hosseini:2017}. In the present paper we have shown that a
similar asymptotics arises with $\sum_k \Delta_k=\tau+\sigma+1$ in a
separate region of parameters, leading to a second blackhole
saddle-point with fugacities that are complex conjugate to those of
the HHZ saddle-point. (See \cite{Benini:2018} for related statements
in the large-$N$ analysis.)

Another conceptual point that we were able to clarify in the special
case with equal charges was the insignificance of the supersymmetric
Casimir energy to the blackhole entropy function in the Cardy-like
limit. Generalizing that discussion to the case with the flavor
fugacities away from the unit circle constitutes another important
open problem related to the present work.

\begin{description}
  \item[Problem 3)] Study the supersymmetric Casimir energy of the $\mathcal{N}=4$ theory with flavor fugacities away from the unit
  circle. In particular, investigate its relevance to the
  blackhole entropy function in the Cardy-like limit.
\end{description}

\subsection*{Note added:} While this work was nearing completion the
preprint \cite{Honda:2019} appeared on arXiv which has some overlap
with our Section~\ref{sec:betaComplex}. Reference~\cite{Honda:2019}
seems to suggest that extra hairy-blackhole
\cite{Markeviciute:2018a,Markeviciute:2018b} saddle-points might
reside on the $W$ wings of the parameter-space. As discussed in
Section~\ref{sec:betaComplex}, we find it more likely that no such
\emph{extra} blackhole saddle-points with entropies as large as the
ones discussed here (as expected to be the case for the hairy
blackholes of \cite{Markeviciute:2018a,Markeviciute:2018b}) exist in
the Cardy-like limit of the index. The existence/interpretation of
extra saddle-points in the large-$N$ analysis \cite{Benini:2018} is
of course a separate issue.

Also, note that \cite{Honda:2019} gives around its Eq.~(2.11) a neat
analytic proof for the $M$-type behavior of the pairwise potential
in a specific subset of the parameter-space for
$\mathrm{arg}\beta>0$ (\emph{i.e.} $\mathrm{Re}(i/\tau\sigma)<0$).
However, the argument below (2.9) there that \emph{the periodicity
of the pairwise potential can be used to extend the analytic proof
of the $M$-type behavior to the whole parameter-space} does not seem
applicable; as illustrated in Figure~\ref{fig:CatastSimp}, in
essentially half of the parameter-space the pairwise potential is in
fact maximized at the origin (this can actually be seen from Honda's
analytic proof in the appropriate region of the parameter-space,
together with the oddity of the potential under
$\Delta_k\to-\Delta_k$). Consequently, the $W$-type behavior of the
pairwise potential in half of the parameter-space when
$\mathrm{arg}\beta>0$, as well as the $M$-type behavior in half of
the parameter-space when $\mathrm{arg}\beta<0$ (and thus also the
second blackhole saddle-point) seem to have been overlooked in
\cite{Honda:2019}.

\begin{acknowledgments}
    It is a pleasure to thank F.~Benini, F.~Bouya, D.~Cassani, S.~Kim,
    P.~Milan, V.~Reys, and especially M.~Hosseini for valuable discussions and correspondences related to this work, as well as J.~Liu and P.~Szepietowski
    for enjoyable collaboration on related projects. I am grateful to F.~Larsen, L.~Pando~Zayas, and M.~M.~Sheikh-Jabbari
    for encouraging me to study related developments, and indebted to F.~Benini and P.~Milan for
    their comments on an earlier draft of this work. All the graphs in this paper were produced by Mathematica.
    This work was supported by the Knut and Alice Wallenberg Foundation under grant Dnr KAW 2015.0083,
    and by the National Elites Foundation of Iran.
\end{acknowledgments}

%
%
%

\end{document}